\documentstyle[aps,prl,epsfig]{revtex}

\begin{document}

\draft 

\title{
Tidal interaction in binary black hole inspiral}

\author{Richard H. Price}
\address{Department of Physics, University of Utah, Salt Lake City, Utah
84112.}
\author{John T. Whelan}
\address{Department of Physical Sciences, University of Texas at Brownsville, 
Brownsville, Texas, 78520.}

\maketitle
\begin{abstract}
\begin{center}
{\bf Abstract}
\end{center}
In rotating viscous fluid stars, tidal torque leads to an exchange of
spin and orbital angular momentum. The horizon of a black hole has an
effective viscosity that is large compared to that of stellar fluids,
and an effective tidal torque may lead to important effects in the
strong field interaction at the endpoint of the inspiral of two
rapidly rotating holes. In the most interesting case both holes are
maximally rotating and all angular momenta (orbital and spins) are
aligned. We point out here that in such a case (i) the transfer of angular
momentum may have an important effect in modifying the gravitational
wave ``chirp'' at the endpoint of inspiral. (ii)~The tidal transfer of
spin energy to orbital energy may increase the amount of energy being
radiated. (iii) Tidal transfer in such systems may provide a
mechanism for shedding excess angular momentum.  We argue that
numerical relativity, the only tool for determining the importance of
tidal torque, should be more specifically focused on binary
configurations with aligned, large, angular momenta.
\end{abstract}

\subsection*{Introduction}\label{sec:intro} 
The problem of the dynamics of a binary pair of comparable mass black
holes (BHs) has been of great interest in connection with sources of
gravitational waves \cite{FlanHughes}.  In the early stages of such a
binary, when the separation of the two holes is large compared to the
size of the holes, the interaction is weak and post-Newtonian
approximations suffice\cite{PN}. At the latest stage, the
``close-limit'' approximation provides the waveform\cite{closelimit}.
Speculations about the strong field interaction have been based on
extrapolating both methods\cite{BD,inspiral}. These methods are
useful, but cannot be trusted to show the effects of phenomena that
are unique to the strong field stage of the binary interaction. Here
we point out that one such effect that has been overlooked may play an
important role in systems with high angular momentum.

The effect can be thought of as the torque exerted on a rapidly
rotating BH due to the tidal distortion created by the other BH. When the
separation $r$ of the BHs is large, and the holes are interacting
weakly, the results of a perturbative analysis, as well as an analogy
with tidal torquing in fluid stars, shows that this tidal torque falls
off as $r^{-6}$ and (as will be shown below) is unimportant.  When $r$
is on the order of the size of the BHs, on the other hand, this
torque may be comparable to the torque exerted by radiation reaction, 
and may therefore be of great importance to the waveform of gravitational 
radiation for just that phase of inspiral that generates the highest
gravitational wave power. 

The late-stage strong-field stage of inspiral, in which tidal torque
may play a crucial role, is not amenable to any presently available
approximation. Reliable answers appear only to be possible with
numerical relativity, solution of Einstein's equations on
supercomputers.  The application of numerical relativity to binary
inspiral has proved to be enormously difficult, but there has been
considerable progress recently\cite{NRref} and an analysis of tidal
torque appears now to be a challenging but not impossible goal. A
major purpose of the present paper is to bring to the attention of the
numerical relativity community the importance of investigating the
phenomenon of late-stage tidal torque.

It is useful to point out that the phrase ``tidal torque'' must not
be taken too literally. The distinction between the spin angular
momentum of orbiting objects, and the orbital angular momentum of
those objects, is meaningful only when the objects are well separated,
or if the individual objects are rigid. These criteria are certainly
not valid at the stage of inspiral at which two disjoint horizons are
merging into a single horizon. Nevertheless, the imagery and vocabulary
of tidal torque, used cautiously, are useful as a background
to approximations and descriptions, and to point to interesting directions
for exploration with numerical relativity. 

When two bodies are in binary orbit, each raises tidal bulges in the
shape of the other. If the objects are rotating, their fluid flows
through this distorted pattern undergoing shear. Viscosity in the
rotating fluid shifts the position of the tidal bulges, and
gravitational forces on the bulges produce a torque.  Bildsten and
Cutler\cite{BildstenCutler} showed that tidal torque could not be
important in inspiral involving neutron star (NS) systems for two
reasons.

 First, the tidal torque becomes significant only when
separation is so small that a NS would have been tidally disrupted.  Their
conclusion that small separation is required is completely consistent
with our findings. Tidal disruption, of course, has no meaning in the
BH case (although the merger of the horizons is somewhat analogous).

The second argument given by Bildsten and Cutler is that tidal torque
for a NS would require unrealistically large coefficient of viscosity,
of order $c\times$(NS radius).  On the basis of perturbative
computations, Hartle\cite{hartle} was the first to suggest that it is
useful to treat horizons as having effective surface shear viscosity
and to be subject to tidal torques in the same manner as fluid
stars. The formalism for ascribing fluid mechanical properties more
generally to highly distorted horizons was subsequently
developed\cite{PT,membrane}. In that formalism the surface shear
viscosity has a value of $1/16\pi$, in the $c=G=1$ units that we
use. If converted to an effective volume viscosity, this meets 
 the Bildsten-Cutler criterion.

\subsection*{Weak field approximation}\label{sec:weak} 
For quantitative, though imperfect, insights into the tidal torquing
process, we present  an analysis here of the effect of tidal torquing in a
BH-BH binary, in the limit of slow rotation and large separation.  In
this approximation, the effect of the tidal field ${\cal E}$ on the
spin angular momentum $S$ of a hole has been shown by Teukolsky to be
\cite{teuk}
\begin{equation}
\frac{dS}{dt}=-\frac{2}{5}\;{\cal E}^2S\mu^3\left[1+3\frac{S^2}{\mu^4}\right]\ ,
\end{equation}
where $\mu$ is the mass of the hole.  The tidal field at one of the
holes, due to the other hole, also of mass $\mu$, at the
the binary separation $r$, is \cite{tidal}
\begin{equation}
{\cal E}=-2\frac{\mu}{r^3}\ .
\end{equation}
These combine to give the tidal torque spindown rate
\begin{equation}\label{dSdt} 
\left.\frac{dS}{dt}\right|_{\rm Tidal}=-\frac{8}{5}\;\frac{\mu^5}{r^6}\;S\left[1+3\frac{S^2}{\mu^4}\right]\
.
\end{equation}
This result must be compared with the rate of radiation of angular
momentum in gravitational waves.  For a quasi-Newtonian circular orbit
of two BHs, each of mass $\mu$, degraded by quadrupole radiation, this
rate is\cite{dEdt}
\begin{equation}\label{dJdt} 
\left.\frac{dJ}{dt}\right|_{GW}=-\frac{256}{5\sqrt{2}}\;\frac{\mu^{9/2}}{r^{7/2}}\ .
\end{equation}
For two extreme ($S=\mu^2$) Kerr holes, the above two equations
predict equal influence of tidal torque and radiation reaction on the
orbital angular momentum $L\equiv J-2S$ for separation $r$ slightly
less than $\mu$. This helps to confirm that tidal torque in BH-BH
binaries, as in NS binaries, can only be important at very small
separations.

The approximations leading to Eqs.\,(\ref{dSdt}) and (\ref{dJdt}),
cannot be trusted to better than rough order of magnitude at small
separations, and hence cannot be used to rule out strong, even
dominant, tidal effects for $r$ several times $\mu$.  Reliable
answers, presumably from numerical relativity, are not yet available.
Lacking those we must make do here with simple illustrative toy models
that show how the importance of tidal torque depends on the unknown
strong-field details.  In particular, we show orbital frequency
$\omega$ as a function of time $t$, for several assumptions about
tidal torque at small $r$.  All models assume that the binary starts
at large separation with $S=\mu^2$, and in all models the Newtonian
expression for orbital angular momentum $L=r\omega^2\mu/2$ is equal to
$J-2S$. The mass $\mu$ of the individual holes is taken to increase
due to tidal dissipation by a law $d\mu/dt=-\omega\,dS/dt$ based on
perturbative models\cite{massincrease}.  The factors shown in the
figure are inserted into the right hand side of Eq.\,(\ref{dSdt}) to
give different models of strong-field tidal torque in comparison with
the Newtonian expression. Each model then consists of the relationship
of (i) $r$, $\mu$ and $\omega$ given by Newtonian theory (ii) the
Newtonian angular momentum relationship $r\omega^2\mu/2=J-2S$, (iii)
the models for $dJ/dt$, $dS/dt$ and $d\mu/dt$ discussed above.

The results in Fig.~1 show what should be expected.  The highest
(solid) curve in Fig.~1 is the evolution of the orbital frequency in
the absence of tidal torque. This curve is unbounded, since the
Newtonian computation on which it is based has no reference to a
special radius at which the inspiral changes its nature. If tidal
torque is included according to Eq.\,(\ref{dSdt}), no change can be
seen in the result until a yet higher frequency is reached. When
$dS/dt$ is enhanced with a factor of $1-6\mu/r$, or its square, the
result is to stall the inspiral near an orbital frequency
$\omega$ equal to the Newtonian value for $r=6\mu$. An analogous
inspiral stall occurs at a lower frequency for enhancement at
$r=8\mu$. These models illustrate that plausibly strong tidal torque
can have profound effects on the waveform. The period of nearly
constant frequency corresponding to the inspiral stall has a
dramatically different character than the frequeny sweep of the
``chirp''\cite{FlanHughes} usually associated with gravitational waves
from binary inspiral. Such a constant frequency signal should be more
easily detectible than a chirp signal.  Furthermore, the stall of the
inspiral means that the binary will be radiating for a prolonged time,
and hence radiating more energy than would be inferred from models
that omit tidal torque. The additional energy radiated is in a sense
the rotational energy of the holes which becomes ``radiatable''
through conversion to orbital motion.

\subsection*{Late stage approximation}\label{sec:close} 

The challenges of numerical relativity mean that we must 
temporarily be satisfied with calculations that are less than
definitive.  The models of the previous section show features of
direct importance to the question of tidal torque, but require guesses
about strong-field interactions.  We next turn to a calculation that
correctly represents strong-field interaction, but is 
related to tidal torque somewhat indirectly.
Figure \ref{threecases} pictures three different initial
configurations of two equal mass holes. In Fig.\,\ref{threecases}a, two
nonspinning holes, at separation $L$, start from rest; in
Fig.\,\ref{threecases}b, two nonspinning holes represent orbital motion
by having initial transverse momentum $P$; Fig.\,\ref{threecases}c
shows two holes, each having spin $S$, starting with no momentum.  A
comparison of Fig.~\ref{threecases}c with the other two cases is
intended to show the following: (i) Since the holes are spinning, the
infall in configuration ($c$) is no longer ``radial,'' as in
Fig.~\ref{threecases}a, and the radiation emitted contains much the
same multipole structure as for the initial configuration in ($b$).
(ii) The addition of spin increases the radiated energy in ($c$) much
the same as transverse motion does in ($b$).

The radiation computation for the configuration in Figs.~2a and 2b
can be computed using the close-limit approximation\cite{closelimit}
for the late stages of inspiral. In that method the spacetime evolving
from the initial data is treated as a perturbation of the spacetime of
the final hole, with the initial separation $L$ taken as the perturbation
parameter.  The actual initial data, i.e.\,, solution of the initial
value problem for Einstein's equations, follow the
Bowen-York\cite{BowenYork} prescription.  This close-limt calculation
was remarkably successful in finding the radiation for the case of
head-on collisions starting with holes at rest, as in Fig.~2a, or
initially moving towards each other. The method has also been
applied\cite{inspiral} with Bowen-York initial data for holes with
transverse, or orbital, momentum $P$ as in Fig.~2b. In that
application it was necessary to treat the momentum $P$, as well as the
initial separation $L$, as a perturbation parameter. The radiated energy
found with 
this multiparameter perturbation calculation is:
\begin{equation}\label{dedtinspiral} 
\frac{\rm Energy}{M}=
9.81\times10^{-5}\left(
\frac{L}{M}
\right)^4
+
1.61\times10^{-2}\left(
\frac{J}{M^2}
\right)^2\ .
\end{equation} 
Here the ADM angular momentum $J$ is given by $J=PL$, the same
expression as in Newtonian theory.  When this result is extrapolated
to conditions approximately representing the ``ISCO,'' the innermost
stable pre-merger circular orbit, the results are in reasonable
agreement (factor of 5 or so) with recent results from numerical
relativity\cite{LazPRL}.

In principle, the configuration in Fig.\,2c, with
Bowen-York\cite{BowenYork} initial data, can be analyzed with $S$ and
$L$ treated as perturbations, but a new complication arises in the
details; the analysis turns out to require higher order perturbation
theory. This analysis is being carried out at present by Gleiser
and Dominguez\cite{reinaldo}, but some features of the result do not
require the completion of that work. In particular, the radiated
energy must have the form
\begin{equation}\label{EbyM} 
\frac{\rm Energy}{M}=
9.81\times10^{-5}\left(
\frac{L}{M}
\right)^4
+
\alpha\left(
\frac{J}{M^2}
\right)^2
\left(
\frac{L}{M}
\right)^4\ ,
\end{equation}
where $J=2S$. Preliminary numerical relativity results by Baker,
Campanelli, Lousto and Takahashi\cite{laz}  suggest that
$\alpha$ is around $2\times10^{-4}$. (This is in accord with our own
order-of-magnitude estimates of $\alpha$.)

For definiteness in a discussion of multipole structures for Fig.~2, we let the
page be the $xy$ plane, so that the angular momenta in Figs.~2b,c are
in the $z$ direction.  The axisymmetric configuration in Fig.~2a can radiate only
in even-parity even-$\ell$ multipoles.  Since the symmetry axis in
Fig.~2a is not the $z$ axis, there will be $m\neq0$ multipoles of the
radiation, but the axisymmetry imposes constraints on the ratio of energy
in multipoles of different $m$.  The less stringent symmetry arguments for the
configurations in Figs.~2b,c require only that the radiation be in
even-parity even-$\ell$ modes or odd-parity odd-$\ell$ modes.  The
multipolar distributions in both Fig.~2b and Fig.~2c greatly differ
from the radial symmetry pattern of Fig.~2a. As an example, much of
the radiated energy in the close-limit calculation for Fig.~2c is in
the odd-parity $\ell=3$ mode (although details require the completion 
of the higher order perturbation theory calculation). This demonstrates
that adding spin to the black holes  changes the 
pattern of the radiated energy, just as orbital motion does.

\subsection*{Discussion}\label{sec:disc} 

We have used the term ``tidal torque'' to refer rather generally to
the processes related to BH spin in the late stage of BH binary
inspiral. We have argued, with oversimplified models, that such
processes may produce crucially important effects that cannot be seen
in post-Newtonian computations, and that have not yet been studied
with numerical relativity. Using a close-limit slow rotation
viewpoint,  we have demonstrated that spin angular momentum and
orbital angular momentum play somewhat the same role in generating
gravitational radiation, and hence that tidal torque (in our generalized
sense) definitely does play an important role if the holes are close 
together. 

The role that tidal torques play in binary inspiral will depend on
whether the holes are ever close enough together. Towards the end of
inspiral the BH-BH system will reach a point at which gradual inspiral
of two bodies is replaced either by a plunge or by the formation of a
single final rotating hole. If the plunge occurs at relatively large
separation, tidal effects may never be significant, and may have no
effect on the chirp waveform. If the two horizons merge to a single
final horizon at sufficiently large separation, the very meaning
of tidal effects is obscured.

The role of tidal effects is closely linked to a long standing
question about binary inspiral. A Kerr black hole, the stationary
solution that is the final state of the inspiral, is constrained to
have its angular momentum and mass related by $J\leq M^2$.  The
angular momentum per unit energy radiated in gravitational waves is
dependent on geometry and frequency. Plausible but nonrigorous
arguments suggest\cite{FlanHughes} that a highly dynamical final hole
will radiate at frequencies too high to reduce $J/M^2$. Similar
arguments make it seem unlikely that $J/M^2$ could be reduced during a
plunge.  An awkward question then arises in the case (as in Fig.~1)
that the individual holes are maximally rotating, and all angular
momentum is parallel. A BH binary just before merger may have excess
angular momentum. This situation may arise through gradual inspiral,
or may be an initial value solution chosen expressly to have excess
angular momentum. Tidal torque provides a way of viewing how the
system can meet the requirements to form a final hole: the transfer
from spin to orbital motion can keep the frequency low (as in Fig.~1)
until $J/M^2$ is sufficiently reduced.

We hope that the points made in this article are sufficiently
persuasive that some effort will be directed to numerically evolving
configurations that would clarify the role of tidal torque. In
particular it would be useful to have studies comparing the evolution
of initial data with and without large spin angular momentum.

\section{Acknowledgment} We gratefully acknowledges the support of the
National Science Foundation to the University of Utah, under grant
PHY-9734871, and to the University of Texas at Brownsville, under
grant PHY-9981795. In addition, we thank John Baker, Manuela
Campanelli, Carlos Lousto and Ryoji Takahashi for making available to
us preliminary numerical results, and to Jorge Pullin and Reinaldo
Gleiser for useful discussions.

%%%%%%%%%%%%%%%%%%%%%%%%%%%%%%%%%%%%%%%%%%%%%%%%%%%%%%%%%%%%

\begin{figure}[h]\label{fig:models} 
\epsfxsize=.45\textwidth\epsfbox{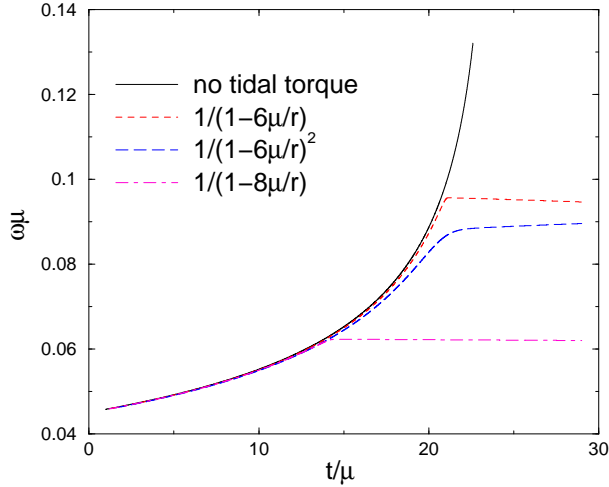}
\caption{Results from simple models of tidal torque. The
Newtonian 
tidal torque $dS/dt$ is multiplied by the factors shown.}
\end{figure}

\begin{figure}[h] 
\begin{center}
\epsfxsize=.28\textwidth\epsfbox{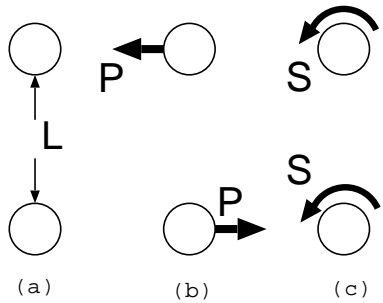}
\end{center}
\caption{Initial conditions for close limit calculations. (a)
Nonspinning holes at rest. (b) Nonspinning holes moving
transversely. (c) Unmoving spinning holes.\label{threecases}}
\end{figure}


\begin{references}
\bibitem{massincrease} See, for example, Ref.\,\cite{membrane}, Sec.\,VII\,B1.
The mass change in the models, in any case, turns out to be unimportant.
\end{references}
\end{document}